# Molecular Dynamics Simulations Study on Ultrathin Cu Nanowires

Jeong Won Kang* and Ho Jung Hwang

Semiconductor Process and Device Laboratory, Department of Electronic Engineering,

Chung-Ang University, 221 HukSuk-Dong, DongJak-Ku, Seoul 156-756, Korea

In order to understand properties of ultrathin copper nanowires, we have simulated several copper nanowires using classical molecular dynamic simulations. As the temperature increases, copper nanowires were transformed into structures of the lowest surface stresses and surface energy, circular cross-section with {111}-like surface. As thickness of copper nanowire increases, the breaking of nanowire and the structural transition hardly occurrs. From studies of angular correlation and radial distribution functions, it was shown that ultrathin {111} nanowires was more stable than that of {100} nanowires. The vibrational frequency of nanowires was different to that of bulk about 3 THz and above 8 THz. The structural properties of cylindrical multi-shell nanowires were greatly different from that of face centered cubic.

*E-mail: kok@semilab3.ee.cau.ac.kr
Tel: 82-2-820-5296
Fax: 82-2-825-1584

# INTRODUCTION

By increasing interests of nanoscience and nanotechnology, nanostructures have been intensively investigated in low-dimensional physics and technological applications in the past decade such as molecular electronic devices, metallic nanowires (NWs), and carbon nanotubes (CNT), etc [1-34]. In recent works for metallic NWs, long metallic NWs with well-defined structures having diameter of several nanometers have been fabricated using different methods. [1-6] Novel helical multi-shell structures were observed in ultrathin gold NWs [1-4], and these have been investigated using molecular dynamics (MD) simulations [7-12]. Many authors have often studied infinite NWs with periodic boundary conditions along the wire axes using MD simulations. Structures of ultra-thin infinite Pb and Al NWs [13, 14], the premelting of infinite Pb NWs [15], the melting of infinite (100) oriented Pt and Ag NWs [16], structures of freestanding Ti NWs [17], and structures of infinite Al, Cu, and Au NWs [12] have been investigated by MD simulations. These results showed that multi-shell and filled metallic NWs exist for several fcc metals. Yanson *et al*. studied multi-shell structures in Na NWs [18]. The stability of Na NWs was studied by modeling them as infinite uniform jellium cylinders and by solving the electronic structures by self-consistent method [19]. Multi-shell NWs were also found for several inorganic layered materials, such as $WS_2$, $MoS_2$, and $NiCl_2$ [20-21]. The strain rate effect induced by amorphization of pure Ni and NiCu alloy NWs has been investigated using MD simulation [23].

Although many studies of metallic NWs have been performed, the present knowledge on properties of metallic NWs is still quite limited. Can unsupported ultrathin Cu NWs maintain their structures at room temperature? Can ultrathin Cu NWs with diameters of several nanometers be found by self-assembly methods? Atomistic simulations have been played important roles in scaling down to nanometer scales and can help in the elucidation of their properties and in the development of new methods for their fabrications and applications. Computational materials science also provides detail microscopic informations on the physical properties of ultrathin metallic NWs. In this work, we have investigated the ulathin freestanding Cu NWs for initial structures, diameters, and temperature.

## SIMULATION METHODS

Interaction between Cu atoms was described by a well-fitted potential function of the second moment approximation of the tight binding (SMA-TB) scheme [35]. The SMA-TB type potential function has been used in atomisitic simulation studies of nanoclusters [36-40] and ultrathin nanowires [17]. This potential is in good agreement with other potentials and with experiment, for bulk [35] and low-dimensional system [33]. The physical values for Cu calculated by the SMA-TB agree with those calculated by other theoretical methods and measured by experiment [33, 36].

MD methods in our previous works [29,36,41] were used, MD timestep was 0.5 fs, MD times were 50 ps, and periodic boundary condition (PBC) was applied along the axes of NWs. In $C_{M4}$ case, MD time was 5 ps. The PBCs of {100}, {111}, and CMS-type Cu NWs were 36.15 Å, 41.738796 Å, 66.51185 Å, respectively.

## SIMULATION RESULTS

We have investigated some structures of Cu NWs by atomistic simulations. Atoms were initially distributed by random positions in cylinders with PBC, and the position of randomly inserted atom has been selected by a criterion, which the nearest-atom-distance is above 2 Å. Figure 1 shows several structures of Cu NWs obtained by different conditions. The total number of atoms in Figs. 1(a) and 1(b) and Figs. 1(c) and 1(d) are $N = 870$ and 480, respectively. The simulated Cu NWs are shown in various structures such as straight-, distorted-, twisted-line, etc. The Cu NW by MD simulation at 300 K is shown in Fig. 1(a), and its cross-section shows the pentagonal structure. The Cu NW in Fig. 1 (b) was obtained by two-step (1) MD simulation at 600 K and (2) the steepest descent method. This case NW was composed of connections of several structures with {111}-like facet. The cross-sections of this NW are similar to polygon, and the left side of this NW is pentagon structure related to decahedron model of Cu nanorods [5]. Figure 1(c) shows the Cu NW obtained by the simulation procedure of Fig. 1(b) case at 300 K. In this case, the Cu NW is a {111}-like structure such as above cases, but its cross-section is rectangular structure.

Figure 1(d) shows the Cu NW obtained by MD simulation at 400 K, and its cross-section is circular structure related to the cylindrical multi-shell (CMS) NWs observed in previous works [3,4,7-14,18]. The structures of CMS-type Cu NWs were explained by semiclassical orbits in a circle [34] and by circular folding of triangular network with orthogonal vector [3,9,34]. Kondo and Takayanagi [3] introduced the notation $n - n'- n''- n'''$ to describe the nanowire consisting of coaxial tubes with $n, n', n'', n'''$ helical atom rows ($n > n' > n'' > n'''$), and structures of CMS-type NWs in Table I are presented by their indexing method. The above results and discussions show that ultrathin Cu NWs could be found in several structures, such as fcc structure, structures related to polyhedral models, and the CMS-type structures with polygonal or circular cross-sections.

In our study of stability for ultrathin Cu NWs, we define initial structures of Cu NWs as Table I. The fcc NWs with rectangular or quasi-circular (polygonal) cross-section shapes and the CMS-type NWs were studied. The label $C_{M4}$ in Table I is not multi-shell structure but composed of triangular network folding, and its structure has been found in simulation works for Al, Pb [13], and Cu [34]. Before MD simulations, initial structures were relaxed by the steepest descent method.

Table II shows MD simulation results whether NWs were broken by the thermal energy, and labels 'O' and 'X' present non-breaking and breaking of NWs, respectively. At 300 K, all NWs except for $C_{M4}$ preserved the original structures and the ultra-thin CMS-type NWs, such as $C_{M6}$ and $C_{M7}$, also stood. The ultra-thin CMS-type Au NWs have been formed in experiments [3]. The ultra-thin CMS-type Cu NWs have been also found in atomistic simulation [34]. Therefore, the CMS-type Cu NWs are expected to composition and observation by experimental methods. As the temperature increases, the broken Cu NWs and the structural transitions by thermal energy are appeared. In the case of $B_{C13}$, the original fcc faces were maintained below 350 K, but some parts of NW partially showed the CMS-type structures between 380 K and 420 K. In this work, the breaking of $C_{M6}$ and $C_{M7}$ were achieved about 455 ± 4 K and 461 ± 5 K, respectively. At 600 K and 900 K, NWs did not maintain the original structures, but cross-sections of NWs were transformed into shapes similar to circle. Table II shows that as the diameter of NW increases, the breaking of NWs do not occur at high temperatures. However, at high temperature MD simulations, {100} NWs were transformed into mixed {111} structures. Figure 2 shows the structures of some NWs at 300 K and $A_{R50}$ NW for temperatures. Figure 2(a) shows the structures of some cases in Table II, and $B_{C13}$ has the small grain-boundary-like structures. At 300 K, since thermal energies of atoms in $A_{R50}$ are low, the structural transition was not founded. However, at 600 K, $A_{R50}$ was transformed into {111} surfaces and its cross-section becomes to hexagon closed to circle with {111}-like surface. At 900 K, the cross-section of $A_{R50}$ was similar to circle with {111}-like surface. In the previous works, metallic NWs with {111}-like surface have been found for the cases of Au, Al, Pb, Cu, Ti, etc. In our previous elongation deformation studies of Cu NWs, some cases of relaxed structures had cylindrical structure with {111}-like surface [29, 33]. As the temperature increases, the freestanding ultrathin metallic NWs were evolved into structures of the lowest surface stresses and surface energy, cylinder forms with {111}-like surface. In MD simulation to investigate structural transition of thicker rectangular {100} NW with 169 atoms/layer, structural transition of NW did not occur below 900 K, but corner rounding of NW was occurred. These results are in agreement with the previous result for Pb NWs [15], that the polygonal cross-sections of NWs were changed into the circular cross-sections at surface melting temperature of NWs.

By investigating the angular correlation function (ACF), the radial distribution function (RDF), and the probability of vibrational frequency (PVF), we discuss structural properties of Cu NWs at room temperature. In Fig. 3, the dashed line is the ACF of the bulk at room temperature. ACFs of NWs started by fcc structures maintain peaks of fcc after 50 ps. Peaks of $B_{C13}$ with {111} planes are sharper than that of $A_{R18}$ and $A_{C18}$ with {100} planes. Therefore, this implies that the structural stability of ultrathin {111} NWs are higher than that of {100} NWs. Since the CMS-type NWs are not general fcc structures but new structure observed in nanostructures, ACFs of the CMS-type NWs are great different from the ACF of the bulk, and do not show 90° peaks but show broad angle distributions from 90° to 120°.

Figure 4 shows RDFs of NWs and the dashed line is the RDF of the bulk at room temperature. Since the CMS-type NWs are not fcc structures, RDFs of the CMS-type NWs do not show the second peaks, related to the second-nearest-neighbor-correlation of fcc, and arrows in Fig. 4 indicate these. The ACF of $A_{R18}$ has broader distribution than ACFs of the other fcc NWs and peaks in the RDF of $A_{R18}$ are hardly distinguishable except for the first peak. The widths of the first peaks of RDFs of NWs except for $B_{C13}$ have wider than that of the bulk. The widths of peaks in the RDF affect the vibrational frequency distribution by interatomic interactions. Figure 5 shows PVFs for NWs and the dashed line is the PVF of the bulk at room temperature. The PVF calculation step is as follows: (1) MD simulations obtain fully relaxed atomic configuration at 300 K. (2) Exerted forces between all interacting atoms are calculated. (3) An atomic position is displaced by 0.03 Å, $\Delta r$, along (1,1,1) vector. Exerted forces between the one atom and its interacting neighbors are calculated. (4) Using the force difference, $\Delta F$, between (2) and (3) steps and $\Delta r$, the force constant, $k$, is calculated by the Hooke's law, $\Delta F = k \Delta r$. (5) The vibrational frequency of harmonic system is $1/2p (k/m)^{1/2}$, where $m$ is atomic mass. After these treatments for all atoms, the probability of the vibrational frequency is calculated. An atomistic interactive cutoff distance is 12 Å. In Fig. 5, PVFs of NWs are different to the PVF of the bulk about 3 THz and above 8 THz. PVF differences above 8 THz are related to interactions between the first nearest neighbor atoms. Since strong interatomic interactions are exerted between the first nearest neighbor atoms, the wider first peak of the RDF is, the higher vibrational frequency has. Since the widths of the first peaks of RDFs of NWs except for $B_{C13}$ have wider than that of the bulk, strong interatomic interactions above 8 THz can be found. In the case of $B_{C13}$ with the narrow first peak of the RDF, the maximum vibrational frequency is about 7.5 THz, which is lower than the maximum vibrational frequency of the bulk. The PVF differences about 3 THz are related to interactions between the second and the third nearest neighbor. Since RDFs for NWs have broader than that of the bulk between the second and the third peaks, PVFs of NWs is higher than the PVF of the bulk about 3 THz.

## SUMMARY

We have investigated structural properties of ultrathin Cu NWs using classical molecular dynamic simulations. This work showed that ultra-thin Cu NWs have several structures, such as fcc, the polyhedral model-related, and the cylindrical multi-shell-type structures with polygonal or circular cross-sections in the different conditions. Unsupported ultrathin Cu NWs have maintained its structure without breaking at room temperature. Cu NWs were transited into structures of the lowest surface stresses and surface energy, circular form with {111} surface, when thermal energy was sufficiently supplied to Cu NWs. As the thickness of Cu NWs increased, the breaking of NW and the structural transition hardly occurred. By investigation of angular correlation function, angle peaks of ultratin Cu NWs with {111} were sharper than that of Cu NWs with {100}, and the structural stability of ultrathin {111} NWs were higher than that of {100} NWs. For the cylindrical multi-shell-type NWs, which are not general fcc structures, angular correlation functions of the cylindrical multi-shell NWs were great different from that of fcc and show broad angle distributions from 90° to 120°. Also, Since the cylindrical multi-shell-type NWs are not fcc structures, radial distribution functions of the cylindrical multi-shell-type NWs did not show the peaks related to the second-nearest-neighbor-correlation of fcc. The probability of vibrational frequency of NWs was different to that of bulk about 3 THz and above 8 THz, and these were related to interactions between first nearest neighbor atoms and related to interactions between the second and the third nearest neighbor, respectively.

Our simulations showed that freestanding ultrathin copper nanowires can maintain their structrures at room temperature and especially, since {111}-like copper nanowires are more stable, it is expected that cylindrical copper nanowires with {111}-like structure could be frequently found in self-assembled ultathin copper nanowires.

Table Caption

Table I. Several ultra-thin Cu nanowires simulated in the work.

Table II. Breaking of ultrathin copper nanowires studied using classical molecular dynamics simulations. Labels 'O' and 'X' present non-breaking and breaking of NWs, respectively.

Figure Captions

Figure 1. Several structures of Cu nanowires with different conditions. (a) a MD simulation at 300 K ($N = 870$). (b) a MD simulation at 600 K and the steepest descent method ($N = 870$). (c) a MD simulation at 300 K and the steepest descent method ($N = 480$). (d) a MD simulation at 400 K ($N = 480$).

Figure 2. Atomic structures of Cu nanowires after MD simulations. (a) Structures of some NWs at 300 K. (b) Structures of $A_{R50}$ for temperature.

Figure 3. Angular correlation functions at 300 K. The dashed line is the angular correlation function in the bulk.

Figure 4. Radial distribution functions at 300 K. The dashed line is the radial distribution function in the bulk.

Figure 5. Probabilities of vibrational frequency at 300 K. The dashed line is the probability of vibrational frequency in the bulk.

TABLES

**Table I.**

|            | Structure       | Atoms/layer | Cross-section shape |
|------------|-----------------|-------------|---------------------|
| $A_{R18}$  | {100}           | 18          | Rectangle           |
| $A_{R50}$  | {100}           | 50          | Rectangle           |
| $A_{R169}$ | {100}           | 169         | Rectangle           |
| $A_{C18}$  | {100}           | 18          | Octagon             |
| $A_{C49}$  | {100}           | 49          | Octagon             |
| $B_{C13}$  | {111}           | 13          | Hexagon             |
| $B_{C54}$  | {111}           | 54          | Hexagon             |
| $C_{M4}$   | CMS-type 4      | 4           | Circle              |
| $C_{M6}$   | CMS-type 5-1    | 6           | Circle              |
| $C_{M7}$   | CMS-type 6-1    | 7           | Circle              |
| $C_{M18}$  | CMS-type 11-6-1 | 18          | Circle              |
| $C_{M34}$  | CMS-type 16-11-6-1 | 34       | Circle              |

**Table II**.

|         | 300 K | 600 K | 900 K |
|---------|-------|-------|-------|
| $A_{R18}$ | O | O | - |
| $A_{R50}$ | O | O | O |
| $A_{C18}$ | O | O | X |
| $A_{C49}$ | O | O | O |
| $B_{C13}$ | O | X | - |
| $B_{C54}$ | O | O | O |
| $C_{M4}$  | X | - | - |
| $C_{M6}$  | O | X | - |
| $C_{M7}$  | O | X | - |
| $C_{M18}$ | O | O | X |
| $C_{M34}$ | O | O | X |

FIGURES

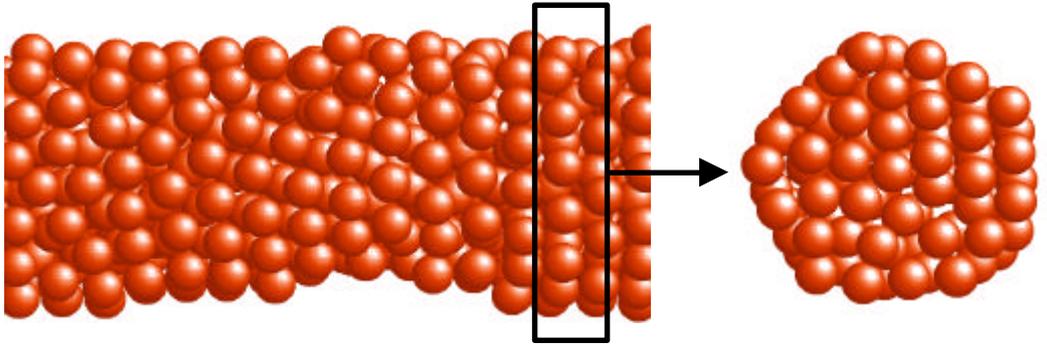

(a)

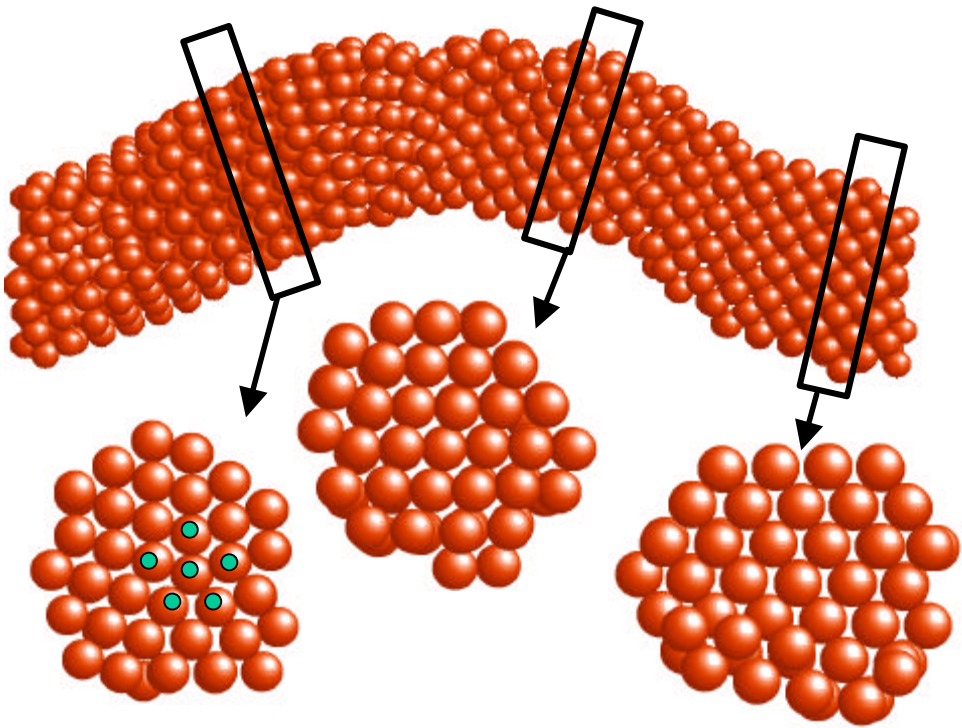

(b)

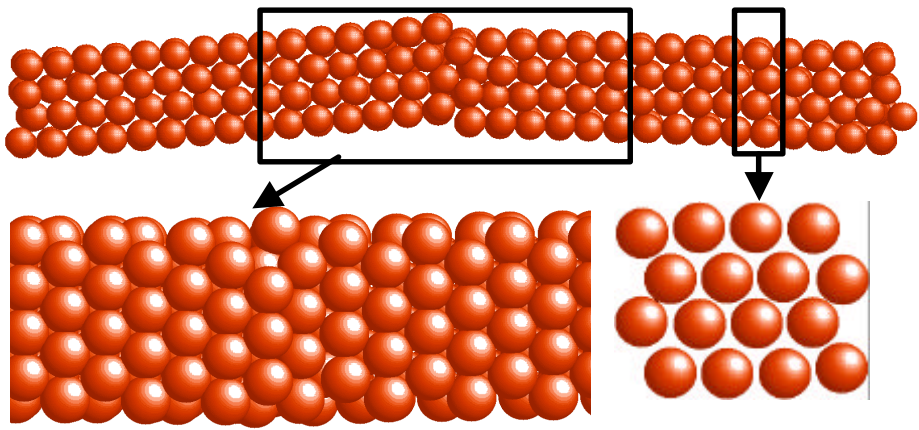

(c)

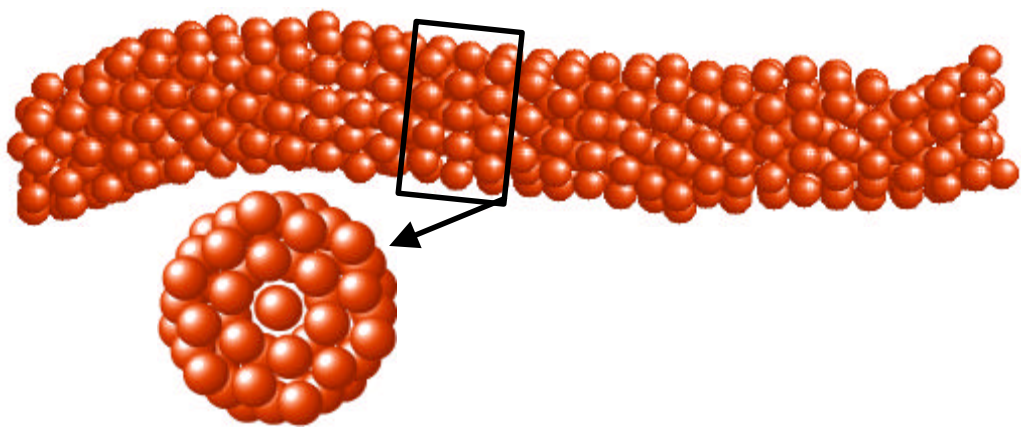

(d)

Figure 1.

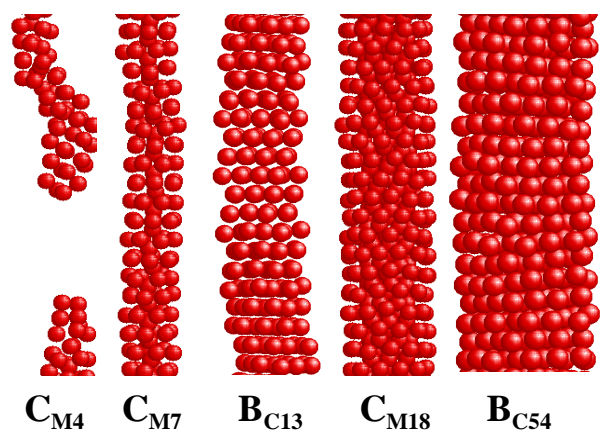

(a)

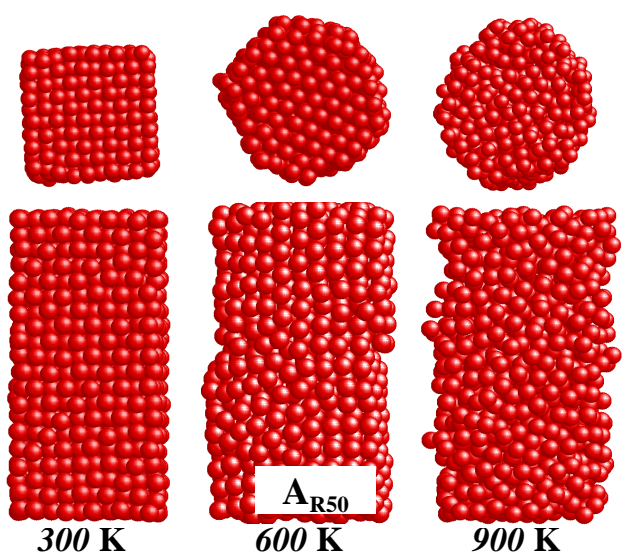

(b)

Figure 2.

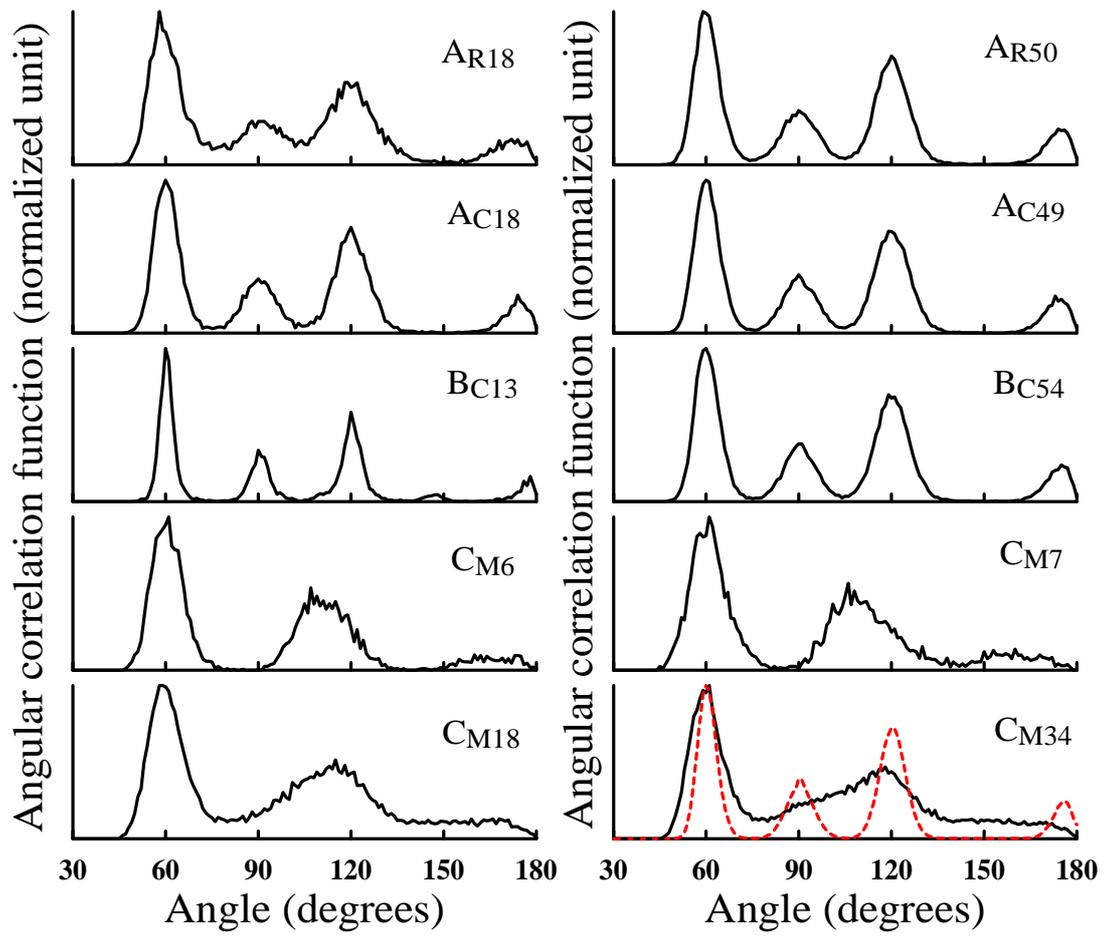

Figure 3.

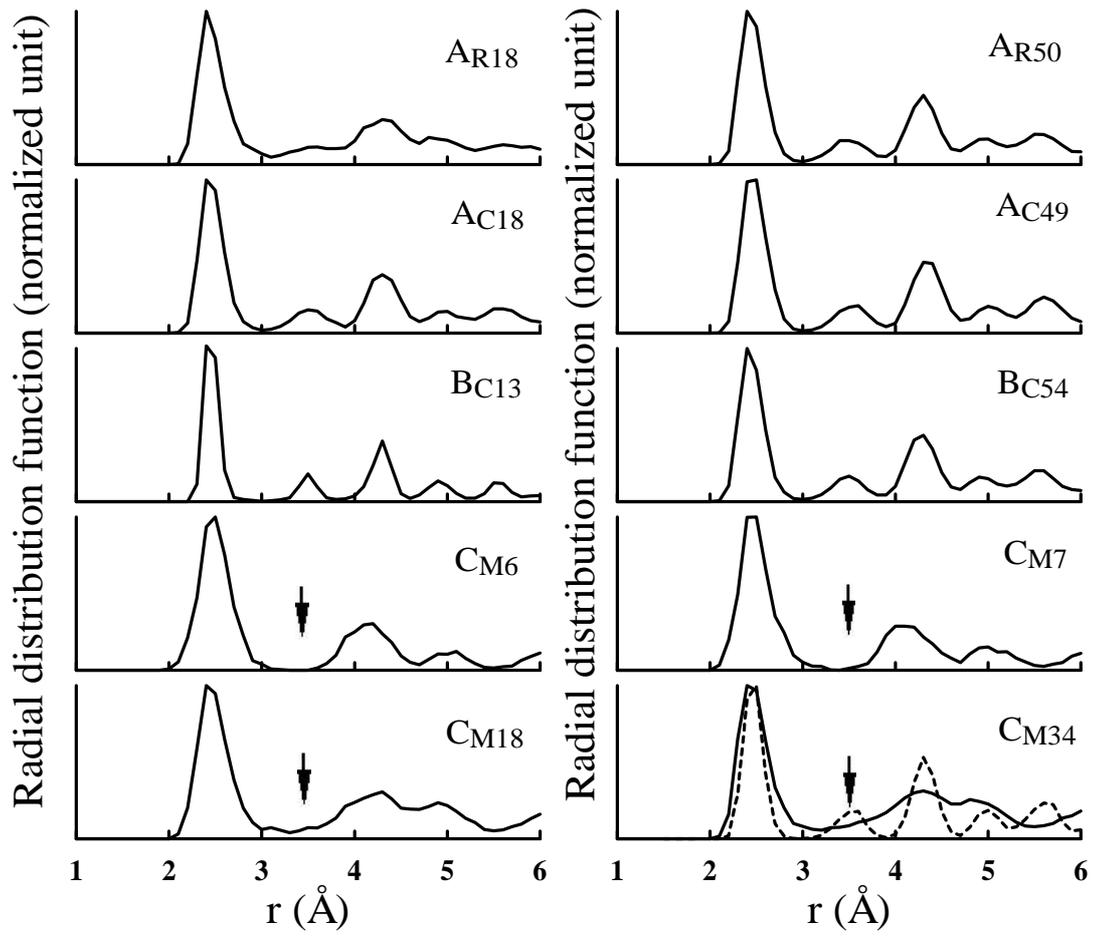

Figure 4.

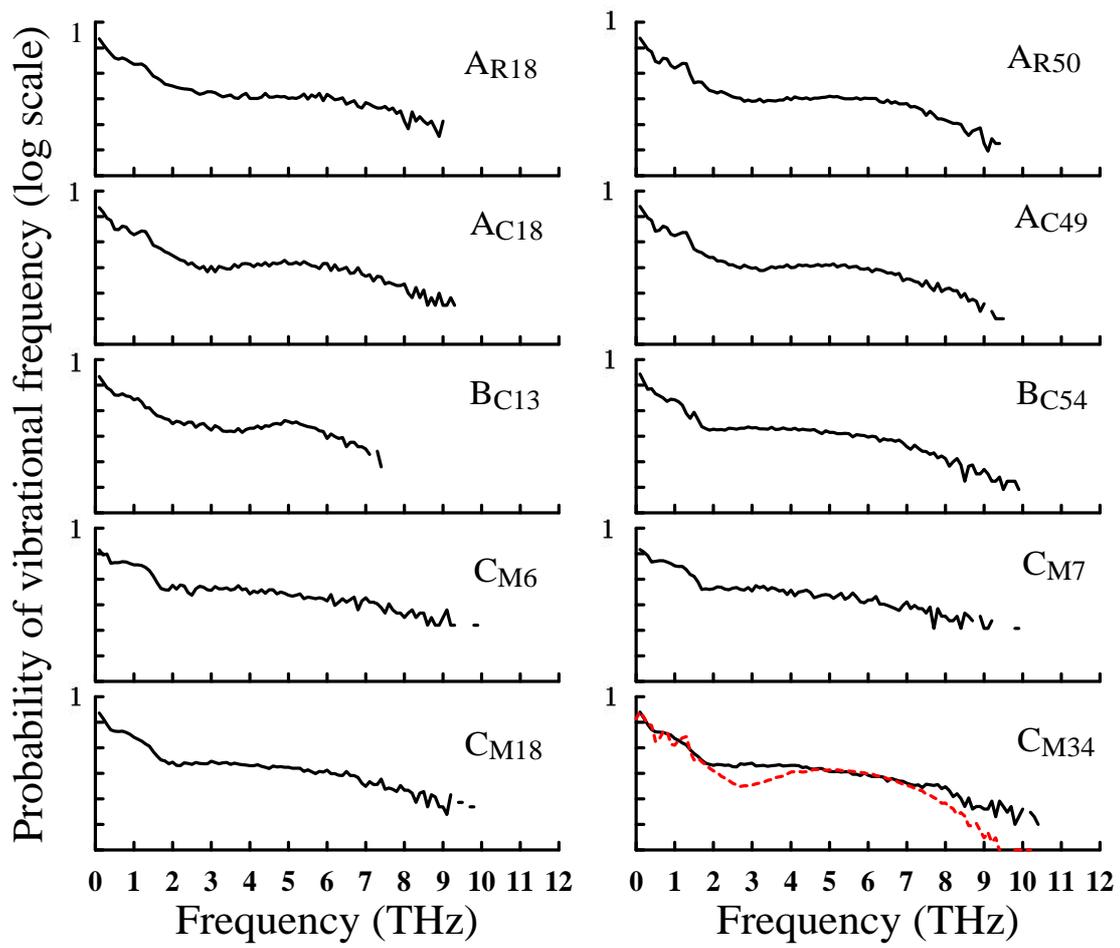

Figure 5.

23